# A2E: Attribute-based Anonymity-Enhanced Authentication for Accessing Driverless Taxi Service

Yanwei Gong, Xiaolin Chang, Jelena Mišić, Vojislav B. Mišić

**Abstract**—Driverless vehicle as a taxi is gaining more attention due to its potential to enhance urban transportation efficiency. However, both unforeseen incidents led by unsupervised physical users' driverless taxi (DT) rides and personalized needs of users when riding in a DT necessitate the authentication of user identity and attributes. Moreover, safeguarding user identity privacy and quickly tracing malicious users if necessary to enhance the adoption of DTs remains a challenge. This paper proposes a novel Attribute-based Anonymity Enhanced (A2E) authentication scheme for users to access DT service. From the security aspect, A2E has attribute verifiability, which is achieved by designing a user attribute credential based on redactable signature. Meanwhile, this attribute credential also satisfies unlinkability and unforgeability. In addition, A2E has enhanced anonymity, which is achieved by designing a decentralized credential issuance mechanism utilizing ring signature and secret sharing, safeguarding user attributes from association with anonymous identities. Moreover, this mechanism provides traceability and non-frameability to users. From the performance aspect, A2E causes low overhead when tracing malicious users and updating credentials. Besides, both scalability and lightweight are satisfied, which contributes to A2E's practicability. We conduct security analysis and performance evaluation to the security and performance capabilities of A2E.

—————————— ◆ ——————————

## 1 INTRODUCTION

Driverless vehicles (DVs) utilizing artificial intelligence [1] have garnered increasing attention for their potential to enhance urban transportation efficiency [2]. Driverless taxis (DTs), as a significant application of DV, are gradually being integrated into transportation systems [3][4]. Due to security is a key factor in determining the further application and promotion of DT [5], numerous studies have focused on assessing critical security mechanisms implemented in DT to ensure secure sensing, positioning, vision, and networking [6][7]. However, the potential of unsupervised physical users' DT rides driven by diverse motivations, such as curiosity, financial gain, or malicious intent, poses unforeseen security risks [8]. Consequently, user authentication, as one of the methods to mitigate these security risks, becomes a crucial requirement among the broader challenges associated with DTs.

Currently, user authentication in DT typically relies on the traditional taxi authentication method, that is, validating the user's mobile phone number linked to the terminal application [9]. However, in the absence of a human driver, such method is not enough for authenticating users who ride DTs because abusing other users' phone numbers is easy.

Moreover, the authentication method may not accommodate personalized user preferences, such as preferences for sharing a DT with individuals of a specific gender, which is particularly relevant for women [10]. Hence, there is a pressing need to develop a novel authentication method for users to access DT service. And there are at least **three challenges**:

i. **Complexity of authentication**. In DT, to ensure the authenticity of user identity so that mitigate malicious behavior, it is essential to authenticate not only user identity but also related identity attributes, such as biometric characteristics. Besides, users accessing personalized and diversified DT services are also required to disclose extra attributes of themselves corresponding to service types [11]. The extra attributes (called service attributes), of which the number is uncertain, still need to be authenticated. Therefore, above problems necessitate the effective implementation of user attributes (related to identity and service) authentication within such dynamic attribute disclosure.

ii. **High requirements for user identity privacy protection**. Since users may choose to disclose their attributes for enjoying personalized DT services, authenticators can know some attributes of users. While user real identity can be protected by conventional anonymization techniques, it still cannot prevent the authenticators from associating specific attributes with the anonymous user, which could facilitate inference attacks [12] to expose the user's real identity. Therefore, how to enhance user identity privacy protection in this context remains a challenge.

• *Yaowei Gong and Xiaolin Chang are with the Beijing Key Laboratory of Security and Privacy in Intelligent Transportation, Beijing Jiaotong University, P.R.China. E-mail: {22110136, xlchang} @ bjtu.edu.cn.*
• *J. Mišić and V. B. Mišić are with Toronto Metropolitan University, Toronto, ON, Canada M5B 2K3. E-mail: {jmisic, vmisic} @torontomu.ca.*

iii. **Traceability with efficiency in user identity privacy preserving**. Since authentication users cannot completely avoid intentional or unintentional malicious users' behaviors, traceability of malicious users is still needed to minimize their impacts. When tracing users, on the one hand, in order to protect privacy, users do not use real identity information to access DT service. On the other hand, the existing traceability methods of [18], [19], [23], [26] require traversing the relevant information of all users, resulting in inefficiency. Thus, how to achieve efficient malicious user traceability in privacy preserving condition is a challenge.

Considering the above challenges, **to the best of our knowledge,** we propose the first *A*ttribute-based *A*nonymity *E*nhanced (A2E) authentication scheme for users when accessing DT service. The unique security and performance features of A2E are listed as follows:

i. **Attribute verifiability with unlinkability and unforgeability**. A2E has attribute verifiability by devising the user attribute credential based on redactable signatures to achieve attribute-based authentication, effectively meeting the intricate authentication demands of user attributes when accessing DT service. Moreover, the credential ensures both unlinkability and unforgeability, guaranteeing the privacy of the user's attribute and the security of the credential, respectively.

ii. **Enhanced Anonymity with Traceability and Non-frameability**. A2E has enhanced anonymity by additionally designing decentralized credential issue mechanism using ring signatures and secret sharing. This ensures that user attributes remain unlinked to the specific anonymous user. Furthermore, traceability and non-frameability are provided by this mechanism, allowing for the identification of malicious users while preventing misuse of user attribute credential.

iii. **Low overhead when tracing user and updating credential**. Compared with [23][26], A2E is low overhead by achieving a notable reduction in the complexity of user traceability, lowering it from $O(A)$ to $O(a)$, where $a$ is substantially smaller than $A$ and both of them are the number of users. Furthermore, A2E is low overhead by devising an efficient credential update mechanism where only part of the credential is modified during updates, thus minimizing the overhead.

iv. **Scalability and lightweight**. A2E is scalable in terms of the minimal additional cost as the number of user disclosed attributes increases during authentication. Moreover, A2E is lightweight in terms of the negligible impact, which is brought by the deployment of itself, on the throughput of achievable DT services.

We conduct formal and informal security analyses, alongside performance evaluations to validate A2E's ability to achieve stated security and performance goals.

The structure of the paper is organized as follows: Section 2 introduces the preliminaries. Section 3 covers the related work and system description. The details of the proposed scheme are elaborated in Section 4. Sections 5 and 6 provide the security and performance analysis of A2E, respectively. Finally, Section 7 concludes the paper.

## 2 PRELIMINARIES

### 2.1 Shamir Secret Sharing

A secret sharing scheme [13] over $Z_q$ consists of a tuple of algorithms $(A_1, A_2)$.

- $A_1(n,t,s)$: This algorithm is used to generate shares $(s_1,...,s_n)$ of the secret $s$. $t$ is the threshold such that $0 < t \leq n$.
- $A_2(s_1,...,s_t)$: This algorithm is used to recover $s$ using the Lagrange's interpolation formula. And $A_2(s_1,...,s_t) = s$.

### 2.2 Redactable Signatures

The redactable signatures proposed in [14] consists of a tuple $(Setup, KeyGen, Sign, Derive, Verify)$ algorithms, which are shown as follows.

- $Setup(1^\lambda)$: This algorithm generates the public parameters according to the security parameter $\lambda$.
- $KeyGen(n)$: This algorithm generates the public-private key pair $(pk, sk)$ according to $n$.
- $Sign(sk, \{m_i\}_{i=1}^n)$: This algorithm generates a signature $\sigma$ for $\{m_i\}_{i=1}^n$ by using $sk$.
- $Derive(pk, \sigma, \{m_i\}_{i=1}^n, \mathcal{D})$: This algorithm generates a signature $\sigma'$ on $\{m_i\}_{i \in \mathcal{D}}$, where $\mathcal{D} \subseteq [1,n]$, according to $(pk, \sigma)$.
- $Verify(pk, \sigma, \{m_i\}_{i=1}^n, \mathcal{D})$: This algorithm returns 1 if $\sigma$ is a valid signature on $\{m_i\}_{i \in \mathcal{D}}$; otherwise returns 0.

### 2.3 Ring Signature

A ring signature scheme [15] is a tuple of algorithms $(GenKey, RSign, RVer)$ which are described as follows:

- $GenKey(1^\lambda)$: This algorithm generates the public-private key pair $(pk, sk)$ according to the security parameter $\lambda$.
- $RSign(PK_{set}, m, sk_i)$: This algorithm generates $\sigma$ for $m$ by using $PK_{set} = \{pk_i\}_{i=1}^N$ and $sk_i$.
- $RVer(PK_{set}, m, \sigma)$: This algorithm returns 1 if $\sigma$ is a valid signature on $m$; otherwise returns 0.

### 2.4 Computational Assumptions

**Discrete Logarithm (DL) assumption.** Given $(g, g^a) \in \mathbb{G}^2$, where $\mathbb{G}$ is a cyclic group of prime order and $g$ is a generator of $\mathbb{G}$, if no efficient adversary can obtain $a$ with non-negligible probability, DL assumption holds.

**Decisional Diffie–Hellman (DDH) assumption** [16]. Assume $(g, g^a, g^b, g^c) \in \mathbb{G}^4$, $\mathbb{G}$ is a cyclic group of prime order and $g$ is a generator of $\mathbb{G}$. If no efficient adversary can distinguish $c = a \cdot b$ from a random element $c \leftarrow Z_p$ with non-negligible probability, DDH assumption holds.

## 3 RELATED WORK AND SYSTEM DESCRIPTION

This section presents the related works and the system description.

### 3.1 Related Work

We summarize the related works about authentication schemes for autonomous vehicles. The characteristics of these schemes were also analyzed in terms of privacy protection and traceability, which is shown in TABLE I.

There have already some scholars focusing on authentication requirements for autonomous vehicles. Bagga et al. [17] presented a protocol for mutual authentication and key agreement in internet of vehicle (IoV)-enabled system. In their proposed protocol, anonymity and untraceability were realized in order to ensure the privacy of vehicles. They also conducted formal security analysis to prove them. Gupta et al. [18] first proposed a rider authentication scheme for DT service. They utilized the touch-signature behavioral biometrics to authenticate users. Both unlinkability and traceability were satisfied. Different from [17], Sagar et al. [19] proposed an authentication method without certificate assistance for IoVs. Besides, data batch verification was realized by using blockchain. They conducted formal analysis and simulation experiment to evaluate the method in terms of security attributes and performance, respectively. Shen et al. [20] designed a lightweight efficient authentication scheme for IoV systems by leveraging blockchain. In [21], for realizing mutual authentication in autonomous IoV network, Adil et al. proposed an access control mechanism by using a double-chain network architecture. The experiment showed that the proposed mechanism was efficient in terms of computation and communication overheads. Li et al. [22] designed an authentication scheme for autonomous truck platooning. In their proposed scheme, zero-knowledge proof was adopted to protect the privacy of truck. Besides, identity verification is implemented by using blockchain. They also conducted the experiments to show the feasibility of the scheme. By using physical unclonable function, Cui et al. [23] designed a multi-factor authentication scheme for autonomous vehicle. The authentication was used to promote a secure remote control of autonomous vehicle. Besides, security and performance of this scheme were superior to other schemes. In addition to [23], Cui et al. [24] also proposed a lightweight authentication for autonomous vehicle. They adopted the blom key to distribute and update session keys efficiently. The security analysis was conducted by using Burrows-Abadi-Needham logic. In [25], an authentication scheme is designed by Xu et al. to realize the access control of autonomous vehicle platoon. Simulations were carried out to show the ultra-reliability and low-latency of the system employed the proposed scheme. Kurt et al. [26] leveraged group authentication to authenticate autonomous vehicles while ensuring their privacy. Sun et al. [27] designed an identity-based broadcasting scheme for IoVs. Their scheme improved both security and efficiency in terms of tamper-resistance and efficient information dissemination, respectively. Besides, security analysis and experiment were conducted to show its satisfaction of the declared goals.

Although there have been some works about authentication for autonomous vehicle, their approaches prioritize vehicle authentication over user authentication and lack implementation of attribute-based authentication. Concurrently, existing schemes failed to safeguard attribute privacy and lack traceability while not with low overhead. To address this gap, this paper introduced A2E to implement attribute-based authentication and anonymity enhancement of user when accessing DT service.

TABLE I COMPARISON OF RELATED WORKS

| Year | Ref | Attribute-based authentication | Anonymity | Traceability | Unlinkability | Non-frameability | Lightweight |
|------|-----|-------------------------------|-----------|--------------|---------------|------------------|-------------|
| 2021 | [17] | × | √ | × | √ | × | × |
| 2022 | [18] | × | × | $O(A)$ [a] | √ | × | × |
| 2022 | [19] | × | √ | $O(A)$ | √ | × | × |
| 2022 | [20] | × | √ | × | × | × | √ |
| 2022 | [21] | × | √ | × | × | × | √ |
| 2023 | [22] | × | √ | × | × | × | √ |
| 2023 | [23] | × | √ | $O(A)$ | × | × | × |
| 2023 | [24] | × | √ | × | × | × | √ |
| 2024 | [25] | × | √ | × | × | × | √ |
| 2024 | [26] | × | √ | $O(A)$ | √ | × | √ |
| 2024 | [27] | × | √ | × | × | × | × |
|      | Ours | √ | Enhanced anonymity | $O(a)$ [b] | √ | √ | √ |

a. $A$ is the number of users.
b. $a$ is also the number of users and substantially smaller than $A$.

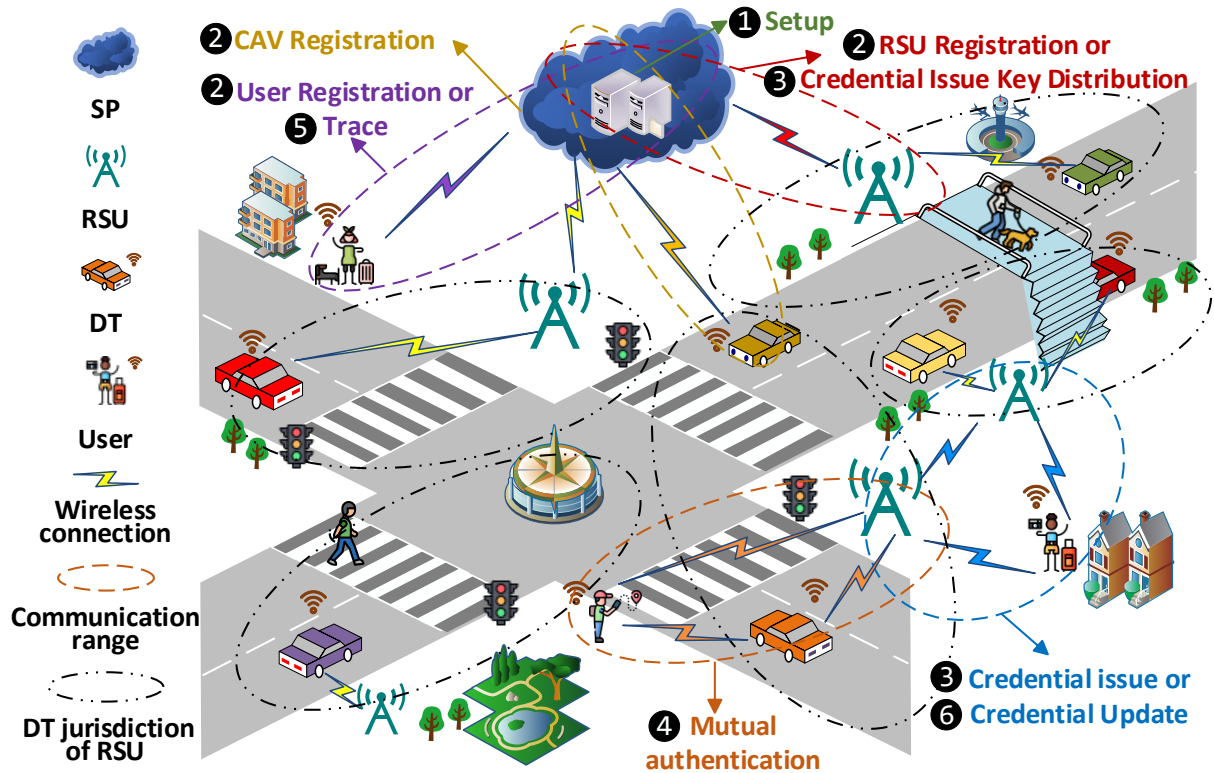

Figure 1. System Framework

### 3.2 System Description

In this section, system description is presented, including system entity, design goals, and security model.

#### 3.2.1 System Entity

In this section, we will introduce entities in the system, which consist of service provider (SP), roadside unit (RSU), driverless taxi (DT), and User. The system framework is shown in Figure 1.

**Service Provider (SP)**. SP is trusted and assumes the responsibility for the registration of RSUs, DTs and Users. It also needs to trace the malicious user in the system. Besides, before User applies for credential, SP is in charge of distributing the credential issue key to RSUs.

**Roadside Unit (RSU)**. RSU is honest-but-curious and can communicate with DTs in its jurisdiction, Users, and other RSUs directly after registration via SP. Before issuing credential for User, each RSU must obtain the credential issue key from SP. Then, when User applies for the credential, each RSU User sends the attribute to will generate a partial credential according to the attribute by using the credential issue key. Some RSUs also need to regenerate partial credentials if one User has attributes updating requirement. Besides, before DT service provision, RSU needs to check User's credential to verify if User meets certain attributes required by this service and then assign DT for this User.

**Driverless Taxi (DT)**. DT is responsible for providing delivery services based on autonomous driving for User directly. It communicates with RSU it belongs to to be assigned which User to serve.

**User**. User can apply services for RSUs when they need to travel or have something to deliver by using its credential. It obtains the credential by sending its attributes to RSUs and aggregates partial credentials returned. When User needs to update its attributes, it only needs to apply partial credentials for the updated attributes instead of re-applying a new credential.

#### 3.2.2 Design Goals

In this section, we present the design goals of A2E in terms of security and performance.

For security, the following security goals should be satisfied.

- **Attribute verifiability**. Attribute verifiability means that RSU can verify attributes which are disclosed by User to access a DT service.
- **Non-frameability**. Non-frameability means that RSUs cannot use User's credential to apply for services, even if the credential is legitimate.
- **Unforgeability**. Unforgeability means that an attacker can't generate valid ring signatures or User credentials.
- **Enhanced Anonymity**. Enhanced Anonymity refers to the fact that even if RSU knows certain attributes of an anonymous User during the mutual authentication, it cannot know which anonymous User those attributes belong to.
- **Traceability**. Traceability means that if a User has misbehaviors, SP can trace to the malicious User.
- **Unlinkability**. Unlinkability means that for the credentials used by Users when applying for services, the attacker cannot link the credentials belonging to the same User together and obtain most of the attributes owned by the User.
- **Resistance of attacks**. A2E should resist the attacks, such as reply attack, man-in-middle attack, and so on.

For performance, the following performance goals should be satisfied.
- **Traceability with Low Overhead**. Traceability with low overhead means that SP does not have to traverse all registered Users' information when tracing malicious Users.
- **Credential Update with Low Overhead**. Credential update with low overhead means that when User wants to update its attributes, only the part of credential corresponding to these attributes needs to update.
- **Scalability**. Scalability means that as the number of attributes disclosed by User in the mutual authentication phase increases, the additional computation and communication overheads are almost negligible.
- **Lightweight**. Lightweight refers to the fact that A2E has little impact on service throughput. That is, when A2E is deployed and when A2E is not deployed, DT can complete almost the same number of services in a certain period.

### 3.2.3 Security Model

In this section, we further define the security model used by the formal security analysis in Section 5 of security goals: unforgeability, anonymity enhancement and unlinkability, which are particularly important for A2E. We first present the **Game Unforgeability**, **Game Anonymity Enhancement** and **Game Unlinkability** used to define the security model for unforgeability, anonymity enhancement, and unlinkability, respectively. All of these games are used to simulate the interactions of entities in A2E and contains many queries which are sent by any probabilistic polynomial time (PPT) adversary $\mathcal{A}$ to the challenger $\mathcal{C}$. Note that the $\{\mathbb{G}_1, \mathbb{G}_2, \text{sy}(pk_{\mathcal{A}_i}, sk_{\mathcal{A}_i}) \text{ mbols}\}$ used in this section are also presented in TABLE II and Section 4.

**Definition 1**. **Unforgeability**. For $\mathcal{A}_\text{I}$, if $Adv_{\mathcal{A}_\text{I}}^{A2E} = |\Pr(win_{\mathcal{A}_\text{I}}^{GUF})| < \varepsilon$ holds where $GUF$ is **Game Unforgeability** and $\varepsilon$ is negligible, then A2E satisfies unforgeability.

**Definition 2**. **Anonymity Enhancement**. For $\mathcal{A}_\text{II}$, if $Adv_{\mathcal{A}_\text{II}}^{A2E} = |\Pr(win_{\mathcal{A}_\text{II}}^{GAE}) - \frac{1}{2}| < \varepsilon$ holds where $GAE$ is **Game Anonymity Enhancement** and $\varepsilon$ is negligible, then A2E satisfies anonymity enhancement.

**Definition 3**. **Unlinkability**. For $\mathcal{A}_\text{III}$, if $Adv_{\mathcal{A}_\text{III}}^{A2E} = |\Pr(win_{\mathcal{A}_\text{III}}^{GUL}) - \frac{1}{2}| < \varepsilon$ holds where $GUL$ is **Game Unlinkability** and $\varepsilon$ is negligible, then A2E satisfies unlinkability.

TABLE II DEFINITION OF SYMBOLS

| Symbol | Description |
| --- | --- |
| $P$ | Pseudorandom generator |
| $(pk_{SP}, sk_{SP})$ | Public-private key pair of SP |
| $ID_{RSU}$ | Real identity of RSU |
| $(pk_{RSU}, sk_{RSU})$ | Public-private key pair of RSU |
| $ID_{DT}$ | Real identity of DT |
| $(pk_{DT}, sk_{DT})$ | Public-private key pair of DT |
| $ID_{User}$ | Real identity of User |
| $(pk_{User}, sk_{User})$ | Public-private key pair of User |
| $(pk_{Issuer}, sk_{Issuer})$ | Public-private key pair of credential issuer |
| $N$ | The number of User attributes |
| $\{attr_j\}_{j=1}^{N}$ | Attributes set of User when applies credential |
| $N'$ | The number of disclosed User attributes |
| $\{attr_j\}_{j=1}^{N'}$ | Attributes set of User when shows credential |
| $cred$ | Credential of User on its all attributes |
| $cred_{\{attr_j\}_{j=1}^{N'}}$ | Credential of User on $\{attr_j\}_{j=1}^{N'}$ |
| $U$ | The number of updated User attributes |
| $sig_X$ | Signature on $X$ |
| $PK_X$ | Public set used to generate $sig_X$ |
| $M$ | The size of public set $PK$ |
| $enc(m)_{pk}$ | Encrypt m by $pk$ using Elliptic Curve Cryptography |
| $dec(enc(m)_{pk})_{sk}$ | Decrypt $enc(m)_{pk}$ by $sk$ using Elliptic Curve Cryptography |
| $sig(m)_{sk}$ | Sign m by $sk$ using Elliptic Curve Cryptography |
| $ver(sig(m)_{sk})_{pk}$ | Verify $sig(m)_{sk}$ by $pk$ using Elliptic Curve Cryptography |
| $\Pr(win_{\mathcal{A}}^{Game})$ | The probability of $\mathcal{A}$ wins the *Game* |

## 4 CONSTRUCTION OF A2E

This section details A2E including six phases, namely, setup, registration, credential issue, mutual authentication, trace, and credential update. The symbols used in A2E are defined in TABLE II.

### 4.1 Setup Phase

In this section, SP will choose the public parameters. The details are as follows:
a) Firstly, SP chooses the security parameter $\lambda$. Additionally, it generates the bilinear pair parameters $(\mathbb{G}_1, \mathbb{G}_2, \mathbb{G}_T, g_1, g_2, e, p)$ and chooses an elliptic curve $E: y^2 = x^3 + ax + b \bmod p$, which also has a cyclic group $\mathbb{G}_E$ with prime order $p$. And $g_E$ is a generator of $\mathbb{G}_E$.
b) Secondly, SP selects a pseudorandom generator $P: \{0,1\}^\lambda \to \{0,1\}^{3\lambda}$. Besides, SP chooses two hash functions: $H_1: \{0,1\}^* \to Z_p, H_2: \{0,1\}^* \to \{0,1\}^{2\lambda}$.
c) Then SP randomly selects the secret key $sk_{SP} \leftarrow Z_p$ and computes its public key $pk_{SP} = sk_{SP} \cdot g_E$.
d) Finally, it keeps $sk_{SP}$ secretly and publishes the

public parameters $\{\mathbb{G}_1, \mathbb{G}_2, \mathbb{G}_T, \mathbb{G}_E, E, g_1, g_2, g_E, e, p, pk_{SP}\}$.

## 4.2 Registration Phase

In this section, RSU, DT and User will register their identity information through SP and obtain the corresponding public-private key pair.

### 4.2.1 RSU Registration

RSU will send its identity information to SP to obtain the public-private key pair. Assume that there are $N$ RSUs, which are denoted as $RSU_i(1 \le i \le N)$. Take $RSU_i$ as for example, the details are as follows.

a) Firstly, $RSU_i$ sends $\{ID_{RSU_i}\}$ to SP.
b) Then SP randomly selects the secret key $sk_{RSU_i} \leftarrow Z_p$ and computes the public key $pk_{RSU_i} = sk_{RSU_i} \cdot g_E$.
c) Finally, SP sends $\{pk_{RSU_i}, sk_{RSU_i}\}$ to the $RSU_i$ via a secure channel.

### 4.2.2 DT Registration

The registration process of DT is same as RSU's described in Section 4.2.1. The difference is that DT obtains its public-private key pair $(pk_{DT}, sk_{DT})$.

### 4.2.3 User Registration

Different from RSU and DT, User needs to obtain the public-private key pair used to generate ring signature for subsequent service requests. The details are as follows:

a) Firstly, User sends $\{ID_{User}\}$ to SP.
b) Secondly, SP receives User messages and generates the public-private key for it. Specifically, SP chooses $\{sk^0_{User,i} \leftarrow \{0,1\}^\lambda, sk^1_{User,i} \leftarrow \{0,1\}^\lambda\}_{i=1}^\lambda$ randomly.
c) Then SP computes $pk_{User,i} = P(sk^0_{User,i}) \oplus P(sk^1_{User,i})$ and $sk_{User,i} = (sk^0_{User,i}, sk^1_{User,i})$. After that, SP obtains the public key $pk_{User} = \{pk_{User,i}\}_{i=1}^\lambda$ and the secret key $sk_{User} = \{sk_{User,i}\}_{i=1}^\lambda$.
d) Finally, SP sends $\{pk_{User}, sk_{User}\}$ to User via a secure channel.

## 4.3 Credential Issue Phase

In this phase, User will apply RSUs for its credential. Before that, RSUs, as issuers, must obtain the credential issue key from SP if they issue a credential for the first time.

### 4.3.1 Credential Issue Key Distribution

The SP will distribute the credential issue key to RSUs, which means each RSU will obtain a secret sharing of the credential issue private key $sk_{issuer}$. Assume that there are $N$ RSUs. The steps are described as follows:

a) Firstly, SP randomly selects $(k_1, k_2) \leftarrow Z_p^2$ and he credential issue private key is $sk_{issuer} = (k_1, k_2)$.
b) Secondly, it computes $K_1 = g_2^{k_1}, K_{2,i} = g_2^{k_2^i}$ ($i \in [1, N]$), and $\tilde{K}_{2,i} = g_1^{k_2^i}$ ($i \in [1, N] \cup [N+2, 2N]$). And the credential issue public key is $pk_{issuer} = (K_1, \{K_{2,i}\}_{i=1}^N, \{\tilde{K}_{2,i}\}_{i=N+2}^{2N})$.
c) Then SP chooses a polynomial $f(x) = k_1 + a_1 x + ... + a_{N-1} x^{N-1}$ with degree $N-1$ and computes [1] $x_i = H_1(ID_{RSU_i}, sk_{SP})$, $f(x_i)$ for $RSU_i(1 \le i \le N)$.
d) Finally, SP sends $\{enc(f(x_i), k_2)_{pk_{RSU_i}}, pk_{issuer}\}$ to $RSU_i$ and publish $\{x_i\}_{i=1}^N$ to $RSU_i$.

When $RSU_i$ obtains the message, it computes $dec(enc(f(x_i), k_2)_{pk_{RSU_i}})_{sk_{RSU_i}}$ to obtain $(f(x_i), k_2)$ and stores $(f(x_i), k_2, pk_{issuer})$.

### 4.3.2 Credential Issue

Before obtaining the credential, User needs to send its attributes to RSUs. In order to avoid the privacy leakage, User signs the attributes by using ring signatures. Let's assume the number of User's attributes is also $N$ and the public set size used by ring signature is $M$. The attributes of User are denoted as $\{attr_j\}_{j=1}^N$ and the public set used to sign $attr_j$ is denoted as $PK_{attr_j} = \{pk_i^{attr_j}\}_{i=1}^M$. The process of credential issue is as follows:

a) Firstly, User chooses $M-1$ random strings $\{str_i^{attr_j} \leftarrow \{0,1\}^\lambda\}_{i=1}^{M-1}$ for the $\{pk_i^{attr_j}\}_{i=1}^M - \{pk_{User}\}$. Additionally, User chooses $\lambda$ random values $\{seed_{i,k}^{attr_j} \leftarrow \{0,1\}^\lambda\}_{k=1}^\lambda$ for each $str_i^{attr_j}$ and computes $r_{i,k}^{attr_j} = (str_i^{attr_j}[k] \cdot pk_{i,k}^{attr_j}) \oplus P(seed_{i,k}^{attr_j})$ for each $seed_{i,k}^{attr_j}$, where $str_i^{attr_j}[k]$ means the $k$-th bit of $str_i^{attr_j}$. After that, User obtains $r_i^{attr_j} = [r_{i,1}^{attr_j}, ..., r_{i,\lambda}^{attr_j}]$ for each $str_i^{attr_j}$.
b) Secondly, User computes $r_{User,k}^{attr_j} = P(sk^0_{User,k})$ for each $\{pk_{User,k}\}_{k=1}^\lambda$. After that, User obtains $r_{User}^{attr_j} = [r_{User,1}^{attr_j}, ..., r_{User,\lambda}^{attr_j}]$.
c) Then User computes $tar_{attr_j} = H_2(\{pk_i^{attr_j}\}_{i=1}^M, \{r_i^{attr_j}\}_{i=1}^M, attr_j)$ and $str_{User}^{attr_j} = str_1^{attr_j} \oplus ... \oplus str_{M-1}^{attr_j} \oplus tar_{attr_j}$. Besides, User computes $seed_{User,k}^{attr_j} = sk_{User,k}^{str_{User}^{attr_j}[k]}$ for each $k \in [\lambda]$.
d) Finally, User obtains $sig_{attr_j} = (PK_{attr_j}, (str_i^{attr_j}[k], seed_{i,k}^{attr_j})_{i \in [M], k \in [\lambda]}, attr_j)$ and sends $\{enc(sig_{attr_j})_{pk_{RSU_j}}\}$ to $RSU_j$.

---

[1] The comma in $H_1$ means concatenation operations for strings, and the same applies to commas in hash functions for the rest of this article.

Once receiving the message, $RSU_i$ will verify the $sig_{attr_j}$ by following steps:

e) Firstly, $RSU_j$ computes $dec(enc(sig_{attr_j})_{pk_{RSU_j}})_{sk_{RSU_j}}$ to obtain the $sig_{attr_j}$. After that, $RSU_j$ parses $sig_{attr_j}$ to obtain $PK_{attr_j}$, $(str_i^{attr_j}[k], seed_{i,k}^{attr_j})_{i \in [M], k \in [\lambda]}$ and $attr_j$.

f) Secondly, $RSU_j$ computes $r_{i,k}^{attr_j} = P(seed_{i,k}^{attr_j})$ if $str_i^{attr_j}[k] = 0$ and else $r_{i,k}^{attr_j} = pk_{i,k} \oplus P(seed_{i,k}^{attr_j})$. After that, $RSU_j$ obtains $r_i^{attr_j} = \{r_{i,k}^{attr_j}\}_{k=1}^{\lambda}$.

g) Then $RSU_j$ computes $tar'_{attr_j} = H_2(\{pk_i^{attr_j}\}_{i=1}^{M}, \{r_i^{attr_j}\}_{i=1}^{M}, attr_j)$. If $tar'_{attr_j} = str_1^{attr_j} \oplus ... \oplus str_M^{attr_j}$ accept $sig_{attr_j}$, else reject.

If $sig_{attr_j}$ is valid, then $RSU_j$ will generate the partial credential for User by following steps:

h) Firstly, $RSU_j$ selects a random number $v_{RSU_j} \leftarrow Z_p$, computes $V_{RSU_j} = g_1^{v_{RSU_j}}$ and sends $\{V_{RSU_j}\}$ to other $\{RSU_i\}_{i=1, i \neq j}^{N}$.

i) Secondly, $RSU_i$ obtains $\{V_{RSU_i}\} \cup \{V_{RSU_j}\}_{j=1, j \neq i}^{N}$. After that, $RSU_i$ computes $cred_{0,i} = \prod_{i=1}^{N} g_1^{v_{RSU_i}}$, $cred_{1,i} = cred_{0,i}^{f(x_i) \prod_{j=1, j \neq i}^{N} \frac{-x_j}{x_i - x_j}}$, and $cred_{2,i} = cred_{0,i}^{k_2^i \cdot attr_i}$.

j) Finally, $RSU_i$ sends $\{cred_{0,i}, cred_{1,i}, cred_{2,i}\}$ to User.

When User obtains $\{cred_{0,i}, cred_{1,i}, cred_{2,i}\}_{i=1}^{N}$, it will aggregate the partial credential by following steps:

k) Firstly, User computes $cred_0 = cred_{0,i}$ and $cred_1 = \prod_{i=1}^{N} cred_{1,i} \cdot cred_{2,i}$.

l) Then, User stores $cred = (cred_0, cred_1)$.

**4.4 Mutual Authentication Phase**

In this phase, User will communicate with $RSU_i$ to obtain DT service. Before that, User needs to show its credential to prove that it meets the corresponding attribute requirements of the service. To ensure that the credential is not abused and the User identity is anonymous, User needs to add a one-time ring signature when displaying the credential. Assume that the public set is denoted as $PK_{req} = \{pk_i^{req}\}_{i=1}^{M}$ and the attributes used in this phase are denoted as $\{attr_j\}_{j=1}^{N'}(N' \leq N)$. Note that the service request is denoted as $req$. Then User will conduct the following steps:

a) Firstly, User randomly selects $(t_1, t_2) \leftarrow Z_p^2$ and computes $cred'_0 = cred_0^{t_1}$, $cred'_1 = cred_1^{t_1} \cdot (cred'_0)^{t_2}$ and $cred' = g_2^{t_2} \cdot \prod_{j=N'+1}^{N} K_{2,j}^{attr_j}$. Besides, it also computes $tar_j = H_1(\{attr_j\}_{j=1}^{N'}, j, cred'_0, cred'_1, cred')$ for each $j \in [1, N']$ and $cred'_2 = \prod_{j=1}^{N'} [\tilde{K}_{2,N+1-j}^{t_2} \cdot \prod_{i=N'+1}^{N} \tilde{K}_{2,N+1+i-j}^{attr_i}]^{tar_j}$. After that, User obtains $cred_{\{attr_j\}_{j=1}^{N'}} = (cred'_0, cred'_1, cred'_2, cred')$ on $\{attr_j\}_{j=1}^{N'}$.

b) Secondly, User chooses $M-1$ random string $\{str_i^{req} \leftarrow \{0,1\}^{\lambda}\}_{i=1}^{M-1}$ for the $\{pk_i^{req}\}_{i=1}^{M} - \{pk_{User}\}$. Additionally, User chooses $\lambda$ random values $\{seed_{i,k}^{req} \leftarrow \{0,1\}^{\lambda}\}_{k=1}^{\lambda}$ for each $str_i^{req}$ and computes $r_{i,k}^{req} = (str_i^{req}[k] \cdot pk_{i,k}^{req}) \oplus P(seed_{i,k}^{req})$ for each $seed_{i,k}^{req}$, where $str_i^{req}[k]$ means the $k$-th bit of $str_i^{req}$. After that, User obtains $r_i^{req} = [r_{i,1}^{req}, ..., r_{i,\lambda}^{req}]$ for each $str_i^{req}$. Then User computes $r_{User,k}^{req} = P(sk_{User,k}^{0})$ for each $\{pk_{User,k}\}_{k=1}^{\lambda}$ and obtains $r_{User}^{req} = [r_{User,1}^{req}, ..., r_{User,\lambda}^{req}]$. Besides, User computes $req = H_2(cred_{\{attr_j\}_{j=1}^{N'}}, T)$, where $T$ is the timestamp, $tar_{req} = H_2(\{pk_i^{req}\}_{i=1}^{M}, \{r_i^{req}\}_{i=1}^{M}, req)$ and $str_{User}^{req} = str_1^{req} \oplus ... \oplus str_{M-1}^{req} \oplus tar_{req}$. Besides, User computes $seed_{User,k}^{req} = sk_{User,k}^{str_{User}^{req}[k]}$ for each $k \in [\lambda]$ and obtains $sig_{req} = (PK_{req}, (str_i^{req}[k], seed_{i,k}^{req})_{i \in [M], k \in [\lambda]}, req)$.

c) Finally, User sends $\{enc(\{attr_j\}_{j=1}^{N'}, cred_{\{attr_j\}_{j=1}^{N'}}, sig_{req})_{pk_{RSU_i}}\}$ to $RSU_i$.

When receiving the request from User, $RSU_i$ will conduct the following steps to authenticate it.

d) Firstly, $RSU_i$ conducts $dec(enc(\{attr_j\}_{j=1}^{N'}, cred_{\{attr_j\}_{j=1}^{N'}}, sig_{req})_{pk_{RSU_i}})_{sk_{RSU_i}}$ to obtain $\{attr_j\}_{j=1}^{N'}$, $cred_{\{attr_j\}_{j=1}^{N'}}$ and $sig_{req}$. After that, it parses $sig_{req}$ to obtain $PK_{req}$, $(str_i^{req}[k], seed_{i,k}^{req})_{i \in [M], k \in [\lambda]}$ and $req$. Then $RSU_i$ computes $\bar{r}_{i,k}^{req} = P(seed_{i,k}^{req})$ if $str_i^{req}[k] = 0$ and else $\bar{r}_{i,k}^{req} = pk_{i,k} \oplus P(seed_{i,k}^{req})$ and obtains $\bar{r}_i^{req} = \{\bar{r}_{i,k}^{req}\}_{k=1}^{\lambda}$. Besides, $RSU_i$ computes $tar'_{req} = H_2(\{pk_i^{req}\}_{i=1}^{M}, \{\bar{r}_i^{req}\}_{i=1}^{M}, req)$ and verify $tar'_{req} = str_1^{req} \oplus ... \oplus str_M^{req}$. If holds, accept $sig_{attr_j}$, else reject.

e) Secondly, $RSU_i$ parses $cred_{\{attr_j\}_{j=1}^{N'}}$ to obtain $(cred'_0, cred'_1, cred'_2, cred')$. Then it computes $tar'_j = H_1(\{attr_j\}_{j=1}^{N'}, j, cred'_0, cred'_1, cred')$ for each $j \in [1, N']$. After that, $RSU_i$ verifies $e(cred'_0, K_1 \cdot cred' \cdot \prod_{j=1}^{N'} K_{2,j}^{attr_j}) = e(cred'_1, g_2)$ and $e(cred'_2, g_2) = e(\prod_{j=1}^{N'} \tilde{K}_{2,N+1-j}^{tar'_j}, cred')$. If both of them hold, $RSU_i$ completes the authentication of User.

f) Finally, $RSU_i$ assigns $DT_i$ within its jurisdiction to the User sends $\{sig(req, ID_{DT_i})_{sk_{RSU_i}}\}$ to User.

User can authenticate $RSU_i$ by conducting $ver(sig(req)_{sk_{RSU_i}})_{pk_{RSU_i}}$. So far, User and $RSU_i$ completes the mutual authentication. Then, User can enjoy the service provided by $DT_i$.

**Remark 1. Correctness**. We will prove that $sig_{req}$ and $cred_{\{attr_j\}_{j=1}^{N'}}$ are valid. Note that $sig_{req}$ is valid when $tar'_{req} = str_1^{req} \oplus ... \oplus str_M^{req}$ holds. We firstly prove the $\overline{r}_i^{req} = r_i^{req}$ holds.

When $pk_i^{req} \neq pk_{User}$, if $str_i^{req}[k] = 0$,

$$\overline{r}_i^{req} = \{P(seed_{i,k}^{req})\}_{k=1}^{\lambda} = \{(0 \cdot pk_{i,k}^{req}) \oplus P(seed_{i,k}^{req})\}_{k=1}^{\lambda}$$
$$= \{(str_i^{req}[k] \cdot pk_{i,k}^{req}) \oplus P(seed_{i,k}^{req})\}_{k=1}^{\lambda},$$
$$= r_i^{req}$$

else

$$\overline{r}_i^{req} = \{pk_{i,k}^{req} \oplus P(seed_{i,k}^{req})\}_{k=1}^{\lambda} = \{(1 \cdot pk_{i,k}^{req}) \oplus P(seed_{i,k}^{req})\}_{k=1}^{\lambda}$$
$$= \{(str_i^{req}[k] \cdot pk_{i,k}^{req}) \oplus P(seed_{i,k}^{req})\}_{k=1}^{\lambda}.$$
$$= r_i^{req}$$

When $pk_i^{req} = pk_{User}$, if $str_i^{req}[k] = 0$

$$\overline{r}_{User}^{req} = \{P(seed_{User,k}^{req})\}_{k=1}^{\lambda} = \{P(sk_{User,k}^{str_{User}^{req}[k]})\}_{k=1}^{\lambda} = \{P(sk_{User,k}^0)\}_{k=1}^{\lambda},$$
$$= r_{User}^{req}$$

else

$$\overline{r}_{User}^{req} = \{pk_{User,k} \oplus P(seed_{User,k}^{req})\}_{k=1}^{\lambda} = \{pk_{User,k} \oplus P(sk_{User,k}^{str_{User}^{req}[k]})\}_{k=1}^{\lambda}$$
$$= \{P(sk_{User,i}^0) \oplus P(sk_{User,i}^1) \oplus P(sk_{User,k}^1)\}_{k=1}^{\lambda}.$$
$$= \{P(sk_{User,i}^0)\}_{k=1}^{\lambda}$$
$$= r_i^{req}$$

Therefore, $\overline{r}_i^{req} = r_i^{req}$ is true. Then

$$tar'_{req} = H(\{pk_i^{req}\}_{i=1}^M, \{\overline{r}_i^{req}\}_{i=1}^M, req) = H(\{pk_i^{req}\}_{i=1}^M, \{r_i^{req}\}_{i=1}^M, req)$$
$$= tar_{req} = (str_1^{req} \oplus str_1^{req}) \oplus ... \oplus tar_{req} \oplus ... \oplus (str_M^{req} \oplus str_M^{req})$$
$$= str_1^{req} \oplus ... \oplus (str_1^{req} \oplus .. \oplus tar_{req} \oplus ... \oplus str_M^{req}) \oplus ... \oplus str_M^{req}.$$
$$= str_1^{req} \oplus ... \oplus str_{User}^{req} \oplus ... \oplus str_M^{req}$$
$$= str_1^{req} \oplus ... \oplus str_M^{req}$$

So far, we prove the validity of $sig_{req}$.

We continue to prove the $cred_{\{attr_j\}_{j=1}^{N'}}$ is valid. In order to get the valid $cred_{\{attr_j\}_{j=1}^{N'}}$, User must get the valid $cred$, which means that $cred_0, cred_1$ is valid. Because $cred_0$ is directly obtained from RSUs and $cred_1$ is obtained by aggregating $\{cred_{1,i} \cdot cred_{2,i}\}_{i=1}^N$. The correct aggregation of $cred_1$ must be ensured, which means that User must recover the $cred_0^{k_1+\sum_{i=1}^N k_2^i \cdot attr_i}$ where $k_1$ is split by secret sharing. The correct aggregation process is as follows:

$$cred_1 = \prod_{i=1}^N cred_{1,i} \cdot cred_{2,i} = \prod_{i=1}^N cred_{0,i}^{f(x_i)\prod_{j=1, j\neq i}^N \frac{-x_j}{x_i-x_j}} \cdot cred_{0,i}^{k_2^i \cdot attr_i}$$
$$= cred_0^{\sum_{i=1}^N f(x_i)\prod_{j=1, j\neq i}^N \frac{-x_j}{x_i-x_j} + k_2^i \cdot attr_i}.$$
$$= cred_0^{k_1+\sum_{i=1}^N k_2^i \cdot attr_i}$$

Since the $cred$ is valid, then the validity of $cred_{\{attr_j\}_{j=1}^{N'}}$ generated by User will be proved. Specifically, we need to prove that $e(cred'_0, K_1 \cdot cred' \cdot \prod_{j=1}^{N'} K_{2,j}^{req}) = e(cred'_1, g_2)$ and $e(cred'_2, g_2) = e(\prod_{j=1}^{N'} \tilde{K}_{2,N+1-j}^{tar_j}, cred')$ hold. The former is proved as follows:

$$e(cred'_1, g_2) = e(cred_1^{t_1} \cdot (cred'_0)^{t_2}, g_2)$$
$$= e((cred_0^{(k_1+\sum_{j=1}^N k_2^j \cdot attr_j)})^{t_1} \cdot (cred'_0)^{t_2}, g_2)$$
$$= e((cred_0)^{t_1 \cdot (k_1+\sum_{j=1}^N k_2^j \cdot attr_j)} \cdot (cred'_0)^{t_2}, g_2)$$
$$= e((cred'_0)^{k_1+\sum_{j=1}^N k_2^j \cdot attr_j} \cdot (cred'_0)^{t_2}, g_2)$$
$$= e((cred'_0)^{k_1+\sum_{j=1}^N k_2^j \cdot attr_j + t_2}, g_2)$$
$$= e(cred'_0, g_2^{k_1+\sum_{j=1}^N k_2^j \cdot attr_j + t_2})$$
$$= e(cred'_0, g_2^{k_1} \cdot g_2^{t_2} \cdot \prod_{j=1}^N K_{2,j}^{attr_j})$$
$$= e(cred'_0, g_2^{k_1} \cdot g_2^{t_2} \cdot \prod_{j=N'+1}^N K_{2,j}^{attr_j} \cdot \prod_{j=1}^{N'} K_{2,j}^{attr_j})$$
$$= e(cred'_0, K_1 \cdot cred' \cdot \prod_{j=1}^{N'} K_{2,j}^{attr_j})$$

The latter is proved as follows:

$$e(cred'_2, g_2) = e(\prod_{j=1}^{N'}[\tilde{K}_{2,N+1-j}^{t_2} \cdot \prod_{i=N'+1}^N \tilde{K}_{2,N+1+i-j}^{attr_i}]^{tar'_j}, g_2)$$
$$= e(\prod_{j=1}^{N'}[\tilde{K}_{2,N+1-j}^{t_2} \cdot \prod_{i=N'+1}^N \tilde{K}_{2,N+1-j}^{k_2^i attr_i}]^{tar'_j}, g_2)$$
$$= e(\prod_{j=1}^{N'}[\tilde{K}_{2,N+1-j}^{t_2+\sum_{i=N'+1}^N k_2^i \cdot attr_i}]^{tar'_j}, g_2)$$
$$= e([\prod_{j=1}^{N'} \tilde{K}_{2,N+1-j}^{tar'_j}]^{t_2+\sum_{i=N'+1}^N k_2^i \cdot attr_i}, g_2)$$
$$= e(\prod_{j=1}^{N'} \tilde{K}_{2,N+1-j}^{tar'_j}, g_2^{t_2+\sum_{i=N'+1}^N k_2^i \cdot attr_i})$$
$$= e(\prod_{j=1}^{N'} \tilde{K}_{2,N+1-j}^{tar'_j}, g_2^{t_2} \cdot \prod_{j=N'+1}^N \tilde{K}_{2,j}^{attr_j})$$
$$= e(\prod_{j=1}^{N'} \tilde{K}_{2,N+1-j}^{tar'_j}, cred')$$

So far, we prove the validity of $cred_{\{attr_j\}_{j=1}^{N'}}$.

**4.5 Trace Phase**

In this process, SP will trace User with malicious behaviors. To be specific, SP will use $(cred_{\{attr_j\}_{j=1}^{N'}}, sig_{req})$ to trace the malicious User by following steps:

a) Firstly, SP parses $PK_{req}$ in $sig_{req}$ to obtain $\{pk_i^{req}\}_{i=1}^M$ and $pk_i^{req} = \{pk_{i,k}^{req}\}_{k=1}^{\lambda}$.

b) Then SP sends $tra$, $PK_{req}$ to $User_i$ who owns $pk_i^{req}$.

Then $User_i$ signs $tra$ in the same as the step b) in Section 4.4 while $tra$ does not need to be recalculated and obtains $sig_{tra}^{User_i}$. Finally, $User_i$ sends $\{sig_{tra}^{User_i}\}$ to SP.

Upon receiving the message, SP will perform the following steps to trace the malicious User:

c) Firstly, SP parses $sig_{tra}^{User_i}$ to obtain $(str_i^{tra}[k], seed_{i,k}^{tra})_{i\in[N], k\in[\lambda]}$.

d) Then for each $pk_{i,k}^{req}(1 \leq k \leq \lambda)$, SP checks if $P(seed_{i,k}^{req}) \oplus P(seed_{i,k}^{tra})_{i\in[M], k\in[\lambda]} = pk_{i,k}^{req}$. If holds, $User_i$ is the malicious User.

Then the User who owns the $pk_i$ is the malicious User. Note that in A2E, SP doesn't need to traverse the information of all registered Users to find the malicious User, leading to efficiency. The detailed analysis is presented in Section 6.

### 4.6 Credential Update Phase

Considering that User may need to update their attributes, generating a new credential for User will bring a heavy burden. Therefore, in our proposed scheme, only the part of credential corresponding to the changed attributes need to update.

Assuming that the updated attributes of User is $\{attr_i\}_{i=1}^{U}$, User firstly also needs to generate the signature for $\{attr_i\}_{i=1}^{U}$ as the step b) in Section 4.4 and obtain $\{sig_i^{attr_i}\}_{i=1}^{U}$. Then, User sends $\{enc(sig_i^{attr_i}, cred_{0,i})_{pk_{RSU_i}}\}$ to $RSU_i$.

When receiving the message, $RSU_i$ firstly conducts $dec(enc(sig_i^{attr_i}, cred_{0,i})_{pk_{RSU_i}})_{sk_{RSU_i}}$ and verifies the $sig_i^{attr_i}$ as the step d) in Section 4.4. Then, $RSU_i$ computes $cred_{2,i} = cred_{0,i}^{k_2 \cdot attr_i}$. Finally, $RSU_i$ sends $\{cred_{2,i}\}$ to User.

After receiving the message, User aggregates the $\{cred_{2,i}\}_{i=1}^{U}$ with other $\{cred_{2,i}\}_{i=U+1}^{N}$ that don't need to update. Firstly, User computes $cred_1 = \prod_{i=1}^{N} cred_{1,i} \cdot cred_{2,i}$. Then, User stores the updated credential $cred = (cred_0, cred_1)$. After that, User completes credential update.

## 5 SECURITY ANALYSIS

In this section, we conduct the security analysis for A2E in two ways: formal and informal. The former is used to prove the security attributes introduced in Section 3.2.3. The latter is used to prove the security goals introduced in Section 3.2.2.

### 5.1 Formal Security Analysis

***Theorem 1***. If, for any PPT $\mathcal{A}_I$, the probability that $\mathcal{A}_I$ can solve the DL is negligible in any polynomial time, then $Adv_{\mathcal{A}_I}^{A2E} = |\Pr(win_{\mathcal{A}_I}^{GUF})| < \varepsilon$ holds and we claim A2E satisfies the unforgeability defined by **Definition 1**.

***Theorem 2***. If, for any PPT $\mathcal{A}_{II}$, $Adv_{\mathcal{A}_{II}}^{A3E} = |\Pr(win_{\mathcal{A}_{II}}^{GAE}) - \frac{1}{2}| < \varepsilon$ holds, then we claim A2E satisfies the anonymity enhancement defined by **Definition 2**.

***Theorem 3***. If, for any PPT $\mathcal{A}_{III}$, the probability that $\mathcal{A}_{III}$ can solve the DL and DDH is negligible in any polynomial time, then $Adv_{\mathcal{A}_{III}}^{A2E} = |\Pr(win_{\mathcal{A}_{III}}^{GUL}) - \frac{1}{2}| < \varepsilon$ holds and we claim A2E satisfies the unlinkability defined by **Definition 3**.

Due to space limitation, the proofs of ***Theorem 1***, ***Theorem 2***, and ***Theorem 3*** are given to the supplemental file.

### 5.2 Informal Security Analysis

In this section, we present the analysis of security goals in an informal way.

**Attribute verifiability**. In the mutual authentication phase, User generates a credential $cred_{\{attr_j\}_{j=1}^{N'}}$ on $\{attr_j\}_{j=1}^{N'}$, signs it with ring signature and sends results $sig_{req}$ to RSU. Then RSU can verify the attributes $\{attr_j\}_{j=1}^{N'}$ of User by verifying $cred_{\{attr_j\}_{j=1}^{N'}}$. Therefore, A2E satisfies attribute verifiability.

**Non-frameability**. Although RSUs can collectively generate a valid User credential $cred$, the credential cannot be used. It is due to that if RSU wants to use $cred$, it must generate a valid ring signature $sig_{req}$ of $cred$. But $sig_{req}$ can be generated only if RSU has User's private key $sk_{User}$, which is unrealistic. Therefore, A2E satisfies non-frameability.

**Unforgeability**. For ring signature, if an attacker wants to generate a valid ring signature, it must obtain $sk_{User}$. For User credential, if an attacker wants to generate a valid User credential, it must obtain $\{f(x_i)\}_{i=1}^{N}, k_2$. Both of them are unrealistic. Therefore, A2E satisfies unforgeability.

**Enhanced Anonymity**. In the mutual authentication phase, although RSU can obtain the attributes $\{attr_j\}_{j=1}^{N'}$ of User, it still doesn't know which anonymous User $\{attr_j\}_{j=1}^{N'}$ belongs to due to the ring signature. Therefore, A2E satisfies enhanced anonymity.

**Traceability**. In the trace phase, SP can trace the malicious User by using $(cred_{\{attr_j\}_{j=1}^{N'}}, sig_{req})$. Then SP can know the true identity of User who generated $sig_{req}$. Therefore, A2E satisfies traceability.

**Unlinkability**. When User credentials are generated, the generated credentials are different even if the same attributes are used due to the random number $(t_1, t_2) \leftarrow Z_p^2$. Thus, even if the attacker gets some User's attributes and corresponding credential, it still cannot determine which attributes and credentials belong to a same User. Therefore, A2E satisfies unlinkability.

**Resistance of attacks**. For replying attack, due to the timestamp, the credential of User can be used only once. For man-in-middle attack, message transmitted in A2E is in ciphertext. Therefore, A2E satisfies resistance of attacks.

## 6 PERFORMANCE ANALYSIS

In this section, we first evaluate the A2E's performance in terms of computation and communication overheads and compare A2E with [23], [26], and [27], which are the latest. Then simulation experiments are conducted to analyze the effect which A2E has on the throughput of achievable DT services. Finally, based on the results obtained, we analyze the performance goals introduced in Section 3.

The experiment setup is as follows. For accessing A2E, JAVA [28] is used to implement it. The cryptographic tool library, JPBC [29], is adopted. BL12 curve is selected to provide a 128 bits security level. And then the size of $\mathbb{G}_E, \mathbb{G}_1, \mathbb{G}_2, \mathbb{G}_T$ and $Z_p$ are 512 bits, 382 bits, 763bits, 4572

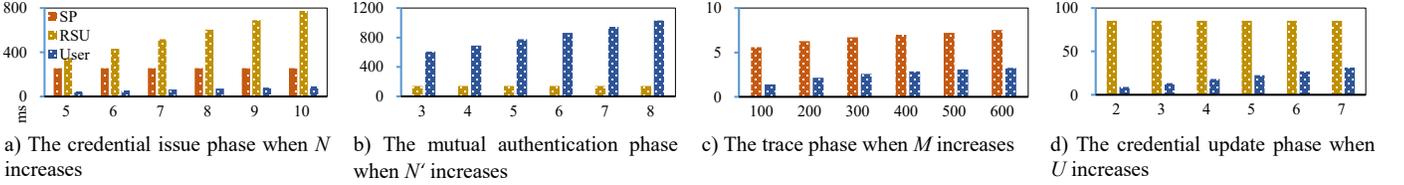

a) The credential issue phase when $N$ increases

b) The mutual authentication phase when $N'$ increases

c) The trace phase when $M$ increases

d) The credential update phase when $U$ increases

Figure 2. The computation overhead of different entities in different phases for different conditions

bits and 256 bits, respectively. We also set $M=100$, $N=5$, $N'=3$ and $U=2$. For the hash function, we use the SHA256 function. Besides, a personal computer with Intel(R)_Core(TM)_i7-10700_CPU@ 2.90GHz 2GB is utilized and the operating system is Ubuntu 20.04.3.

**6.1 Computation Overhead Analysis**

In this section, we evaluate the computation overhead of A2E according to the parameter setting introduced. The computation overhead of basic operations is firstly tested and showed in 错误!未找到引用源。.

For the secret shared shard operation used to compute $f(x_i)$ according to $x_i$, we compute the time cost when $f(\cdot)$ is in degree $5$. Note that the degree is related to the number of User's attributes. To be specific, they are equal. $T_{CGME}$ and $T_{ISME}$ mean the modular exponentiation operation the element in $\mathbb{G}_1$ or $\mathbb{G}_2$, and $Z_p$, respectively.

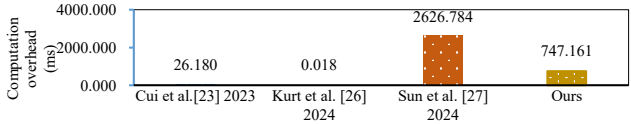

Figure 3. The comparison of computation overhead of related works in the mutual authentication phase

According to the 错误!未找到引用源。, we summarize the computation overhead of each entity in different phases except the setup phase in A2E, of which results are presented in 错误!未找到引用源。. When analyzing computation overhead of each phase, the time cost of parsing string is ignored. For the credential issue phase, when RSU verifies the signature from User, it needs to compute $r_{i,k}^{attr_j} = P(seed_{i,k}^{attr_j})$ if $str_i^{attr_j}[k]=0$ and else $r_{i,k}^{attr_j} = pk_{i,k} \oplus$ $P(seed_{i,k}^{attr_j})$. Due to the uncertainty of $str_i^{attr_j}$, we compute $O_{XOR} = \sum_{i=1}^{1000}\sum_{k=1}^{128} str_i^{attr_j}[k]$ to analyze the time cost of RSU in credential issue phase where $O_{XOR}$ represented the number of times required for the XOR operation, This also applied to calculate the time consumed by RSU to verify User signatures during the mutual authentication phase. For the trace phase, $P(seed_{i,k}^{req}) \oplus P(seed_{i,k}^{tra})_{i \in [M], k \in [\lambda]} = pk_{i,k}^{req}$ needs to be calculated before tracing to the malicious User. Therefore, we assume that the number of public keys needed to be retrieved before tracing to the malicious User is half the size of the public key set, i.e. $2/M=50$. As can be seen from 错误!未找到引用源。, compared with SP and RSU, the computation overheads of User are smaller, which also makes A2E more practical.

We also evaluate the impact on the computation overheads for different entities in different phases as $N$, $N'$, $M$ and $U$ increase. As can be seen from Figure 2.a), with the growth of $N$, the computation overheads of SP and User in the credential issue phase grow slowly, while RSU's increases significantly. In Figure 2.b), the computation overhead of RSU in the mutual authentication phase hardly changes as $N'$ increases. The User's is linearly related to the growth of $N'$. Considering that there is a large gap between the computation overhead of SP and User in the trace phase, Figure 2.c) shows the computation overhead of them in logarithmic form. As can be seen from it, with the growth of $M$, the calculation cost of SP increases rapidly, but it is still in the acceptable range. Figure 2.d) shows the trend of computing overhead for RSU and User in the credential update phase as $U$ increases. It can be seen that the calculation cost of RSU is independent of $U$.

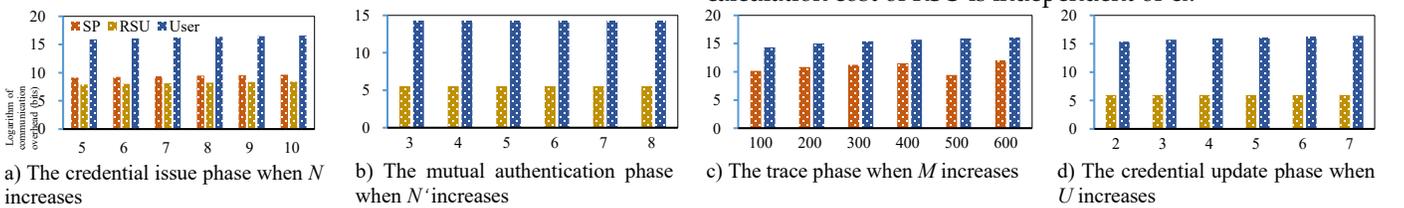

a) The credential issue phase when $N$ increases

b) The mutual authentication phase when $N'$ increases

c) The trace phase when $M$ increases

d) The credential update phase when $U$ increases

Figure 4. The communication overhead of different entities in different phases for different conditions

In Figure 3, We compare A2E with other related works in terms of the computation overhead in the mutual authentication phase. Compared with [23] and [26], the computation overhead of A2E is significantly higher, which is caused by the ring signature used to enhance the anonymity of User. Even so, the computational overhead of our scheme is significantly lower than that of [27].

**6.2 Communication Overhead Analysis**

The communication overhead of A2E is presented in this section, which is also according to the parameter set introduced, and 错误!未找到引用源。 shows the results about communication overhead of basic elements.

According to the 错误!未找到引用源。, we also summarize the computation overhead of each entity in different phases except the setup phase in A2E, of which results are presented in 错误!未找到引用源。. Note that due to the huge size of the public set used to generate a ring signature, in practice, communication overhead can be significantly reduced by transferring the User identity set instead of *PK*. And the receiver can obtain the User public key according the User identity in order to verify the ring signature. User can also choose the smaller *PK*. In addition, although the communication overhead of the credential issue phase is high, User only needs to apply for the credential once. Besides, although the communication overhead is high as can be seen from 错误!未找到引用源。, 6G technology can provide powerful communication capabilities in internet of vehicle [30]. Therefore, whether the communication overhead is large or not has little impact on the practical application.

The communication overheads for different entities at different phases as *N*, *N′*, *M* and *U* increase are evaluated of which results are shown in Figure 4. Due to the existence of the ring signature in A2E, the communication overhead between different entities varies greatly. Therefore, we show the logarithm of each entity's communication overhead in Figure 4. It can be seen from Figure 4.a) that the increase of *N* has little impact on the communication overheads of SP, RSU and User in the credential issue phase, but User's is higher than other entities', which also occurs in Figure 4.b), c), and d),. This is because the User needs to generate the ring signature and send the corresponding set of public keys. In practice, it can choose to send only the User identity set corresponding to *PK*, which greatly reduces the communication overhead of itself. Besides, Users can choose a smaller set of public keys according to the actual communication capability to further reduce the communication overhead. In Figure 4.b), as *N′* grows, the communication overhead of RSU is unchanged in the mutual authentication phase, and User's is growing slowly. As we can see from Figure 4.c), the growth of *M* has a significant impact on the communication overhead for both SP and User in the trace phase. This is because SP needs to communicate with the User set corresponding to *PK* in order to trace the malicious User and User in the User set needs to generate the ring signature using *PK*. Figure 4.d) shows that when *U* grows, the communication overhead of RSU has no change, implying the independence between RSU's and *U*. The communication overhead of User increases with the increase of *U*. Because each additional attribute that needs to be updated requires User to communicate with an additional RSU.

Figure 5 shows the comparison of communication overhead of related works in the mutual authentication phase. Due to the large gap of communication overhead of them, we also present the results in the logarithmic form. It can be seen that the communication overhead of A2E is higher than [23], [26] and [27], which is also due to the usage of the ring signature. However, the communication overhead caused by ring signature can be greatly reduced in practice by sending User identity set instead of *PK*.

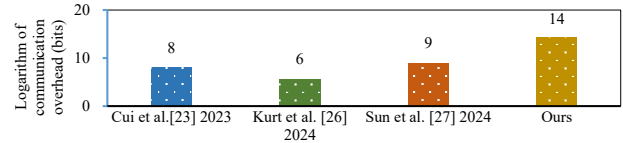

Figure 5. The comparison of communication overhead of related works in the mutual authentication phase

## 7 Conclusion

This paper introduces an *A*ttribute-based and *A*nonymity *E*nhanced (A2E) authentication scheme for user when accessing driverless taxis service. A2E designs user attribute credentials and a decentralized credential issuance mechanism to realize attribute-based authentication and anonymity enhancement, respectively. Then we make comprehensive security analysis and performance evaluations to validate A2E's security and performance goals. In the future work, we are committed to further improving the A2E's performance to enhance its practicality.

**Yanwei Gong** is currently pursuing the Ph.D. Degree in Cyberspace Security at Beijing Key Laboratory of Security and Privacy in Intelligent Transportation, Beijing Jiaotong University, China. His interests include identity authentication protocol related to MEC and fully homomorphic encryption acceleration.

**Xiaolin Chang** is currently a professor at the School of Computer and Information Technology, Beijing Jiaotong University, China. Her current research interests include Edge/Cloud computing, Network security, security and privacy in machine learning. She is a senior member of IEEE.

**Jelena Mišić** is Professor of Computer Science at Toronto Metropolitan University, Ontario, Canada. She serves on editorial boards of IEEE Trans. Veh. Technol., Comput. Netw, Ad hoc Netw, Secur. Commun. Netw, Ad Hoc Sens Wirel Netw, Int. J Sens Netw, and Int. J of Telemed Appl. She is a Fellow of IEEE and Member of ACM.

**Vojislav B. Mišić** is Professor of Computer Science at Toronto Metropolitan University, Ontario, Canada. His research interests include performance evaluation of wireless networks and systems and software engineering. He serves on the editorial boards of IEEE Trans. Cloud Comput., Ad hoc Netw, Peer Peer Netw Appl, and Int. J Parallel, Emergent Distrib. Syst.. He is a Senior Member of IEEE and Member of ACM